\documentclass[a4paper,11pt]{article}

\usepackage{enumitem}
\usepackage{float}
\usepackage[margin=2.5cm]{geometry}
\usepackage{graphicx}
\usepackage[labelfont=bf,labelsep=period]{caption}
\usepackage{times}
\usepackage{url}
\usepackage{xspace}
\usepackage{mathrsfs}
\setlist[]{topsep=2pt,partopsep=2pt,parsep=2pt,itemsep=2pt}
\usepackage{algorithmicx}
\usepackage{algpseudocode}
\usepackage[section]{algorithm}

\floatstyle{ruled}
\newfloat{algo}{htbp}{algo}
\floatname{algo}{Algorithm}

\usepackage{amsmath} %
\allowdisplaybreaks[2]          %

\usepackage{amssymb} %

\usepackage{amsthm} %
\newtheoremstyle{plain-boldhead}%
  {\topsep}%
  {\topsep}%
  {\itshape}%
  {}%
  {\bfseries}%
  {.}%
  { }%
  {\thmname{#1}\thmnumber{ #2}\thmnote{ (\bfseries #3)}}%
\newtheoremstyle{definition-boldhead}%
  {\topsep}%
  {\topsep}%
  {\normalfont}%
  {}%
  {\bfseries}%
  {.}%
  { }%
  {\thmname{#1}\thmnumber{ #2}\thmnote{ (\bfseries #3)}}%
\theoremstyle{plain-boldhead}
\newtheorem{theorem}{Theorem}
\newtheorem{proposition}[theorem]{Proposition}
\newtheorem{Lemma}[theorem]{Lemma}
\newtheorem{corollary}[theorem]{Corollary}

\theoremstyle{definition-boldhead}
\newtheorem{definition}[theorem]{Definition}
\newtheorem{remark}[theorem]{Remark}
\newtheorem{example}[theorem]{Example}

\floatstyle{ruled}
\newfloat{algo}{htbp}{algo}
\floatname{algo}{Algorithm}

\DeclareMathOperator{\lcm}{lcm}

\begin{document}

\title{\bf An Algebraic Model For Quorum Systems}

\author{Alex Pellegrini$^1$\\
  University of Bern\\
  \url{alex.pellegrini@inf.unibe.ch}
  \and Luca Zanolini$^1$\\
University of Bern\\
  \url{luca.zanolini@inf.unibe.ch}
}

\date{}

\maketitle

\footnotetext[1]{Institute of Computer Science, University of Bern,
  Neubr\"{u}ckstrasse 10, 3012 CH-Bern, Switzerland.}

\begin{abstract}
Quorum systems are a key mathematical abstraction in distributed fault-tolerant computing for capturing trust assumptions. A quorum system is a collection of subsets of all processes, called quorums, with the property that each pair of quorums have a non-empty intersection. They can be found at the core of many reliable distributed systems, such as cloud computing platforms, distributed storage systems and blockchains.
In this paper we give a new interpretation of quorum systems, starting with classical majority-based quorum systems and extending this to Byzantine quorum systems. We propose an algebraic representation of the theory underlying quorum systems making use of multivariate polynomial ideals, incorporating properties of these systems, and studying their algebraic varieties. To achieve this goal we will exploit properties of Boolean Gr\"obner bases. The nice nature of Boolean Gr\"obner bases allows us to avoid part of the combinatorial computations required to check consistency and availability of quorum systems. Our results provide a novel approach to test quorum systems properties from both algebraic and algorithmic perspectives.
\end{abstract}

\section{Introduction} 
Quorum systems are a key mathematical abstraction in distributed fault-tolerant computing for capturing trust assumptions.
Quorums help in reaching higher availability and fault-tolerance in distributed systems \cite{DBLP:journals/eatcs/Vukolic10}.
From a classical point of view, a quorum system is a collection of subsets of all processes $\mathcal{P}$, called quorums, with the property that each pair of quorums have a non-empty intersection. It is a generalization of the concept of a \emph{majority} in a democratically organized group and it is used to ensure consistency in the context of crash failures, i.e., when processes stop executing steps \cite{DBLP:books/daglib/0025983}. 
Given a fail-prone system $\mathcal{F} \subseteq 2^{\mathcal{P}}$, that is, a collection of subsets containing all processes that may at most fail together in some execution, we say that $\mathcal{Q} \subseteq 2^{\mathcal{P}}$ is a \emph{quorum system} with respect to $\mathcal{F}$ if each pair of elements of $\mathcal{Q}$ has non empty intersection and for every element $F$ of $\mathcal{F}$ there exists an element of $\mathcal{Q}$ that does not intersect $F$.
However, if the failing processes can deviate in any conceivable way from their algorithm, the above definition is not useful. Malkhi and Reiter \cite{DBLP:journals/dc/MalkhiR98} introduced a generalization of classical quorum systems called \emph{Byzantine quorum systems}, strengthening the definition in a way that the pair-wise intersection contains also some correct processes. To do this, they present the so-called \emph{dissemination quorum systems} and \emph{masking quorum systems}.

We study the properties of quorum systems through Boolean multivariate polynomial ideals encompassing their properties. These structures admit generating bases called Gr\"obner bases. These allow for efficient ways to compute solutions of the set of polynomials of the original ideal. We study quorum properties by inspecting such solutions sets, called ideal varieties.

Though Gr\"obner bases were originally introduced by Buchberger in polynomial
rings over fields \cite{Buc65}, many works have used Gr\"obner
bases over coefficient rings that are not fields. Among them,
Gr\"obner bases of Boolean polynomial rings (Boolean Gr\"obner bases) introduced in
\cite{sakai1988boolean,sakai1991boolean} have appealing properties. For a comprehensive description of Boolean polynomial rings and Boolean Gr\"obner Basis the reader is referred to \cite{DBLP:journals/jsc/SatoISNS11}.
In this work we exploit Boolean Gr\"obner basis according to the \textsc{lex} monomial ordering. This allows us to analyze the structure of the elimination ideals starting from the ones we define. Moreover, given an ideal, using its Gr\"obner basis we can retrieve useful information regarding its associated variety. For instance, the size of the variety and the size of the set of standard monomials of the Gr\"obner basis have a correlation.

The remainder of this work is structured as follows. Section~\ref{sec:2} introduces preliminaries and relevant definitions. We give an overview of the algebraic concepts that we need during the paper. In Section~\ref{sec:3} we build an algebraic model in a way that the remaining part of the paper becomes independent from set-theoretic operations. In Section~\ref{sec:4} we show our results in applying the algebraic model first to classical quorum systems and then to Byzantine ones. In particular, we consider the Byzantine dissemination and masking quorum systems. Finally, Section~\ref{sec:algo} presents algorithms that exploit our representation for testing if set systems are usable as quorum systems and in distributed computing protocols.

\section{Preliminaries and notation}
\label{sec:2}

Our results will heavily rely on the algebraic background that we will introduce in this section. Many of the definitions and proofs can be found in \protect{\cite[Section~4.3]{DBLP:books/daglib/0091062}} and \cite{DBLP:conf/ascm/SatoNI07,gao2009counting,DBLP:journals/corr/abs-1104-0746}, we are hereby giving the result without proofs.

\begin{definition}[Ideal]
Let $(R,\cdot,+)$ be a ring and $(R,+)$ its additive group. A subset $I \subseteq R$ is called an \emph{ideal} if
\begin{itemize}
\item $(I,+)$ is a subgroup of $(R,+)$;
\item For every $r \in R$ and $f \in I$ we have that $r\cdot f \in I$.
\end{itemize}
If $f_1,\ldots, f_s$ live in the ring $\mathbb{K}[X_1,\ldots, X_n]$ over a field $\mathbb{K}$, then $\langle f_1,\ldots , f_s\rangle$ is an ideal of $\mathbb{K}[X_1,\ldots, X_n]$.
We will call $\langle f_1,\ldots, f_s\rangle$ the ideal \emph{generated} by $f_1,\ldots, f_s$.
\end{definition}

\begin{definition}[Boolean ring]
    A commutative ring $\mathbb{B}$ with identity $1$ is called a \emph{Boolean ring} if every element $a \in \mathbb{B}$ is idempotent, i.e. $a^2 = a$.
    Let $\mathbb{B}$ be a Boolean ring, the quotient ring $\mathbb{B}[X_1,\ldots,X_n]/\langle X_1^2 -X_1,\ldots,X_n^2-X_n \rangle$ is a Boolean ring. It is called \emph{Boolean polynomial ring} and denoted by $\mathbb{B}(X_1,\ldots,X_n)$.
\end{definition}

From now on we will consider $\mathbb{B}=\mathbb{F}_2$, which is actually a field, and work with the polynomial ring $\mathbb{B}(X_1,\ldots,X_n)$ unless otherwise specified. 
Hilbert \cite{Hilbert1970} proved that a polynomial ring over a Noetherian ring is Noetherian. This means, in our case, that every ideal in $\mathbb{B}(X_1,\ldots,X_n)$ admits a finite basis.
Notice that a polynomial in $\mathbb{B}(X_1,\ldots,X_n)$ is uniquely represented by a polynomial of $\mathbb{B}[X_1,\ldots,X_n]$ that has at most degree 1 for each variable $X_i$. Sets of variables such as $\{X_1,\ldots,X_n\}$, $\{Y_1,\ldots,Y_n\}$ and $\{T_1,\ldots,T_n\}$ are abbreviated  by $\bar{X}, \bar{Y}$ and $\bar{T}$, respectively.
With small letters like $p,q$ we will usually denote $n$-tuples of elements of $\mathbb{B}$ for some $n$.
Let $f\in \mathbb(\bar{X},\bar{Y})$ be a polynomial and pick $p \in \mathbb{B}^n$, then $f(p,\bar{Y})$ denotes a polynomial in $\mathbb{B}(\bar{Y})$ obtained by specializing $\bar{X}$ with $p$.

\begin{definition}[Sum and product of ideals]
    If $I$ and $J$ are ideals of a polynomial ring $\mathbb{B}[X_1,\ldots,X_n]$ then the \emph{sum} of $I$ and $J$, denoted as $I+J$, is the set
    \begin{equation}
        I + J = \lbrace f + g ~|~ f\in I\text{ and } g\in J \rbrace
    \end{equation}
    and their \emph{product}, denoted $I \cdot J$, is defined to be the set
    \begin{equation}
        I \cdot J = \lbrace f\cdot g~|~f\in I \text{ and } g \in J  \rbrace.
    \end{equation}
\end{definition}

\begin{definition}[Variety]
Let $I \subseteq \mathbb{B}(\bar{X})$ be an ideal. Define the \emph{variety} of $I$ as the set
\begin{equation}
\mathscr{V}(I) = \lbrace x\in  \mathbb{B}^{n} ~|~ f(x)=0 \quad \forall f\in I\rbrace.
\end{equation}
     An ideal $I \in \mathbb{B}[X_1,\ldots,X_n]$ is \emph{0-dimensional} if the associated variety $\mathscr{V}(I)$ is a finite set, i.e. $\#\mathscr{V}(I) < \infty$.
\end{definition}

\begin{theorem}
    If $I$ and $J$ are ideals in $\mathbb{B}[X_1,\ldots,X_n]$, then $\mathscr{V}(I+J) = \mathscr{V}(I) \cap \mathscr{V}(J)$ and $\mathscr{V}(I \cdot J) = \mathscr{V}(I) \cup \mathscr{V}(J)$.
\end{theorem}

We rely on the following theorems to support one of our main results in Section~\ref{sec:4}.
During this work we will use the set of variables denoted by $\bar{X},\bar{Y},\bar{T}$ and $\bar{Z}$ with respect to a block order $\bar{T} < \bar{Z} < \bar{Y} < \bar{X}$. Unless otherwise specified we use the \textsc{lex} monomial ordering, i.e. $\prec = \prec_{\textsc{lex}}$ on variables in each block, e.g. $X_n \prec X_{n-1} \prec \cdots \prec X_1$. Recall that \textsc{lex} is an elimination ordering on the monomials \protect{\cite[Definition~11]{Shibuta}}.

Assume, for example, that we are working on the Boolean polynomial ring defined on blocks of variables $\bar{X},\bar{Y}$ and $\bar{Z}$ with block and elimination order as mentioned before. Given an ideal $I$ and block $\bar{X}$, from now on we will denote with $\pi_{\bar{X}}(\mathscr{V}(I))$ the natural projection on the $n$ coordinates corresponding to the $\bar{X}$ block.

\begin{example}
Let $I \in \mathbb{B}(\bar{X},\bar{Y},\bar{Z})$ with \textsc{lex} order. Let $v=(v_1,\ldots,v_{3n}) \in \mathscr{V}(I)$, then $\pi_{\bar{Y}}(v) = (v_{n+1},\ldots,v_{2n})$. 
\qed

\end{example}

\begin{definition}[Elimination ideal]
Given $I = \langle f_1, \ldots , f_s\rangle \subseteq \mathbb{B}[X_1, \ldots , X_n]$, the \emph{$l$-th elimination ideal $I_l$} is
the ideal of $\mathbb{B}[X_{l+1},\ldots, X_n]$ defined by
\begin{equation}
    I_l = I \cap \mathbb{B}[X_{l+1}, \ldots , X_n].
\end{equation}
\end{definition}

Important results concerning elimination ideals and related varieties are the ``Elimination Theorem''  and ``Extension Theorem'' \protect{\cite[Theorem~2 and 3]{DBLP:books/daglib/0091062}}, respectively. We will only leverage on the latter. Loosely speaking, it says that, given a point in the variety of the elimination ideal, there exist at least one extension of that point that lies in the variety of the original ideal.
In Boolean rings we can define a special case of the Extension Theorem.

\begin{theorem}[Boolean Extension Theorem]
\label{th:extension}
    Let $I$ be a finitely generated ideal in a Boolean polynomial ring $\mathbb{B}(X_1,\ldots,X_n,Y_1,\ldots,Y_n)$. For any $p \in \mathscr{V}(I \cap \mathbb{B}(\bar{X}))$ there exists $q \in \mathbb{B}^n$ such that $(p,q) \in \mathscr{V}(I)$.
\end{theorem}

 Moreover, Gao \cite{DBLP:journals/corr/abs-1104-0746} proved a result over general finite fields $\mathbb{F}$ which relates the variety of an elimination ideal, with the corresponding projection of the variety of the original ideal. We give hereby a specific case of the theorem, restricted to our environment. The proof that this form of the theorem holds is straightforward. 
 
\begin{theorem}[\protect{\cite[Theorem~3.1]{DBLP:journals/corr/abs-1104-0746}}]\label{th:projvariety}
Let $I \subseteq \mathbb{B}(\bar{X}, \bar{Y})$ be an ideal. Then
\begin{equation}
    \pi_{\bar{X}}(\mathscr{V} (I)) = \mathscr{V}(I \cap \mathbb{B}(\bar{X})).
\end{equation}
\end{theorem}

Let $\prec$ be a monomial ordering on $\mathbb{B}(X_1,\ldots, X_n)$. Then we define the \emph{leading monomial} $LM(f) = \max_\prec\{X^\alpha ~|~ X^\alpha \in f\}$ and the \emph{trailing monomial} $TM(f) = \min_\prec\{X^\alpha ~|~ X^\alpha \in f\}$.

We give here a brief introduction to Gr\"obner basis theory and related results we will need in later sections. We will mostly exploit such results in order to design and support algorithms in Section~\ref{sec:algo}.

\begin{definition}[Gr\"obner basis]
    Let $I \subseteq \mathbb{B}[\bar{X}]$ be an ideal and $\mathcal{G} = \{ g_1,\ldots, g_t\} \subseteq \mathbb{B}[\bar{X}]$ a set of polynomials such that $I = \langle \mathcal{G} \rangle$. We say that $\mathcal{G}$ is a \emph{Gr\"obner basis} for $I$ if and only if
    \begin{equation}
        \forall f \in I, f \neq 0, \quad \exists~g_i \in \mathcal{G}\text{ s.t. }LM(g_i)|LM(f).
    \end{equation}
\end{definition}

Usually, a Gr\"obner basis of a set of polynomials is computed using either a variant of Buchberger's algorithm \cite{Buc65} or using Faugere's F4 \cite{FAUGERE199961} or F5 \cite{Faug2} algorithm. 

In some of our results we will only need to know the size of the variety of an ideal. This can be done without computing the variety at all. We will leverage on what follows to conclude our theses.

Let $I \in \mathbb{B}(\bar{X})$ be an ideal. We consider the set of leading monomials of $I$ defined as $LM(I) = \{LM(f)~|~ f \in I\}$. We can define the monomial ideal generated by $LM(I)$ as the polynomial ideal $\langle LM(I) \rangle \in \mathbb{B}(\bar{X})$. Define also

\begin{definition}[Standard monomials]
The set of standard monomials of any ideal $I \in \mathbb{B}(\bar{X})$ is denoted as follows.
\begin{equation}
    SM(I) = \{X_1^{\alpha_1}\cdots X_n^{\alpha_n} ~|~ X_1^{\alpha_1}\cdots X_n^{\alpha_n} \not \in \langle LM(I) \rangle, \alpha_i \in \mathbb{B}\}.
\end{equation}
When an ideal $I$ has a Gr\"obner basis $\mathcal{G}$, we also write the standard monomial set of $I$ as $SM(\mathcal{G})$, and call it the standard monomial set of $\mathcal{G}$.
\end{definition}

The following result allows us to avoid the computation of varieties in some of our results. We give a specification related to our environment.
\begin{theorem}[\protect{\cite[Theorem~3.2.4]{gao2009counting}}]\label{th:standard}
    Let $I \in \mathbb{B}[\bar{X}]$ be a 0-dimensional ideal and $\mathcal{G}$ a Gr\"obner basis for $I$. Then
    \begin{equation}
        |\mathscr{V}(I)| = |SM(\mathcal{G})|.
    \end{equation}
\end{theorem}
 
\section{Set abstraction}
\label{sec:3}

In this section, we construct a model for representing set operations algebraically.

Let $n\in \mathbb{N}$, $\mathcal{P} = \lbrace P_1,\ldots,P_n\rbrace$ be any set and $S \subseteq \mathcal{P}$, say $S=\lbrace P_{j} : j \in \mathcal{J}\rbrace$ with $\mathcal{J} \subseteq \{1,\ldots,n\}$ a set of indexes. Define $\varphi:2^{\mathcal{P}} \rightarrow  \mathbb{B}^n$ as follows 
\begin{equation}
\varphi(S) \mapsto \sum_{j \in \mathcal{J}} \textbf{e}_{j}
\end{equation} 
where $\{\textbf{e}_i\}_{i=1,\ldots,n}$ is the canonical basis of $ \mathbb{B}^n$ and we consider the usual vector sum on $\left( \mathbb{B}^n, +\right)$. With $\varphi(S)_i$, we denote the $i$-th coordinate of the vector $\varphi(S)$. Moreover, for $\mathcal{A} \subseteq 2^\mathcal{P}$ we set $\varphi(\mathcal{A}) = \{ \varphi(A) ~|~ A \in \mathcal{A}\}$.
Define the inverse $\varphi^{-1}:  \mathbb{B}^n \rightarrow 2^\mathcal{P}$ transforming a vector in $ \mathbb{B}^n$ into the associated set in $\mathcal{P}$
\begin{equation}
\varphi^{-1}(\sum_{j \in \mathcal{J}} \textbf{e}_{j}) \mapsto \{P_j : j \in \mathcal{J}\}.
\end{equation}

\begin{example}
Let $n = 5$ then $\mathcal{P}=\lbrace P_1,\ldots, P_5\rbrace$, let also $S = \{P_1,P_3,P_4\}$ with $\mathcal{I}=\{1,3,4\}$ then
\begin{equation}
\varphi(S) = \textbf{e}_1 + \textbf{e}_3 + \textbf{e}_4 = (1,0,1,1,0).
\end{equation}
Given $(1,0,1,1,0)$, then 
\begin{equation}
\varphi^{-1}((1,0,1,1,0)) = \{P_1, P_3, P_4\} = S.
\end{equation}
\qed
\end{example}

\begin{remark}
We write $q = \sum_{i=1}^n c_i\textbf{e}_i$ with $c_i \in \mathbb{B}$ for every $i=1,\ldots,n$. 
We denote the support of a vector $q \in  \mathbb{B}^n$ as $\mathrm{Supp}(q) = \{i ~|~ c_i=1\}$.
\end{remark}
With this notation we can represent the powerset of $\mathcal{P}$ in terms of the vector space~$\mathbb{B}^n$. 

In the rest of this section we define a set of Boolean multivariate polynomials that we will use to define the polynomial ideals encompassing quorum properties. This part of the section, in other words, translates set-theoretic operations into a polynomial representation.

Consider the two sets of variables $\bar{X} = \{X_1,\ldots,X_n\}$ and $\bar{Y} = \{Y_1,\ldots,Y_n\}$ and define $\gamma \in \mathbb{B}(\bar{X},\bar{Y}) = \mathbb{B}(X_1,\ldots X_n,Y_1,\ldots Y_n)$ as
    \begin{equation}\label{eq:pt-poly}
\gamma(\bar{X},\bar{Y}) = \prod_{i=1}^n (X_iY_i + Y_i + 1) \quad+1.
\end{equation}

The next lemma states that a zero of $\gamma$ is an element $v \in \mathbb{B}^{2n}$ such that the support of the first half contains the support of the second half.

\begin{Lemma}\label{lem:support}
Let $q,p \in  \mathbb{B}^n$ then $\mathrm{Supp}(q) \subseteq \mathrm{Supp}(p)$ if and only if 
\begin{equation}
\gamma(p,q) = 0.
\end{equation}
Let $P,Q \subseteq \mathcal{P}$. Then $Q \subseteq P$ if and only if 
\begin{equation}
\gamma(\varphi(P),\varphi(Q)) = 0.
\end{equation}
\end{Lemma}
\begin{proof}
    Notice that $\gamma(\bar{X}, \bar{Y}) = 0$ if and only if $(X_i + 1)Y_i = 0$ for every $i = 1,\ldots, n$. This is equivalent to say $(X_i,Y_i) \neq (0,1)$ for every  $i = 1,\ldots, n$.
    For the second part of the claim just set $p = \varphi(P)$ and $q = \varphi(Q)$.
\end{proof}

In other words, we have that $\gamma$ reflects the set inclusion operator.

Define $\sigma \in \mathbb{B}(\bar{X},\bar{Y})$ as 
\begin{equation}\label{eq:sigma}
    \sigma(\bar{X}, \bar{Y}) = \prod_{i=1}^n (X_iY_i +1).
\end{equation}

Equation (\ref{eq:sigma}) defines a polynomial whose zeros are vectors of the form $v \in \mathbb{B}^2$ such that the supports of the two components intersect. 

\begin{Lemma}\label{lem:supportint}
Let $q,p \in  \mathbb{B}^n$ then $\mathrm{Supp}(p) \cap \mathrm{Supp}(q)$ if and only if 
\begin{equation}
\sigma(p,q) = 0.
\end{equation}
Thus, given $P,Q \subseteq \mathcal{P}$. Then $P \cap Q \neq \emptyset$ if and only if 
\begin{equation}
     \sigma(\varphi(P),\varphi(Q)) = 0.
\end{equation}
\end{Lemma}
\begin{proof}
    By construction $\sigma(\bar{X}, \bar{Y}) = 0$ if and only if there exists $i \in \{1,\ldots,n\}$ such that $X_iY_i = 1$ which is equivalent to say that $X_i = Y_i = 1$.
    For the second part of the claim just set $p = \varphi(P)$ and $q = \varphi(Q)$.
\end{proof}
Similarly to the case of $\gamma$, the next corollary is a translation of Lemma~\ref{lem:supportint} to sets, through the application of $\varphi$. In other words, $\sigma$ reflects the set intersection operation.

Finally, we define $\delta \in \mathbb{B}(\bar{X},\bar{Y},\bar{T})$ as
\begin{equation}\label{eq:delta}
\delta(\bar{X},\bar{Y},\bar{T}) =\prod_{i = 1}^n(T_iX_iY_i +X_iY_i + 1).
\end{equation}

Equation (\ref{eq:delta}) defines a polynomial which will reflect a special set operation whose explanation and proof is the goal of the next lemma and corollary. We will need such an operation when it comes to talk about dissemination quorum systems.

\begin{Lemma}\label{lem:supportdissemin}
Let $p,q,r \in  \mathbb{B}^n$ then $(\mathrm{Supp}(p) \cap \mathrm{Supp}(q)) \not\subseteq \mathrm{Supp}(r)$ if and only if 
\begin{equation}
\delta(p,q,r) = 0.
\end{equation}
Thus, given $P,Q,R \subseteq \mathcal{P}$. Then $(P \cap Q) \not\subseteq R$ if and only if 
\begin{equation}
     \delta(\varphi(P),\varphi(Q),\varphi(R)) = 0.
\end{equation}
\end{Lemma}
\begin{proof}
Assume first $(\mathrm{Supp}(p) \cap \mathrm{Supp}(q)) \not\subseteq \mathrm{Supp}(r)$. Thus there exists an $i$ such that $p_i=q_i=1$ and $r_i=0$. So, $r_ip_iq_i+p_iq_i+1=0$, and it easy to show that in any other case, it has value $1$. It follows that
\begin{equation}
\delta(p,q,r) = 0
\end{equation}
On the other hand, assume $\delta(p,q,r) = 0$. Assume $(\mathrm{Supp}(p) \cap \mathrm{Supp}(q)) \subseteq \mathrm{Supp}(r)$, meaning that for every $i$ such that $p_i=q_i=1$, also $r_i=1$. Then $r_ip_iq_i+p_iq_i+1=1$ for each of those $i$. We obtain $\delta(p,q,r) = 1$ which is a contradiction. Observe that if $\mathrm{Supp}(p) \cap \mathrm{Supp}(q) = \emptyset$, it means that for every $i=1 \ldots n$, $p_i \neq q_i$. Then $r_ip_iq_i+p_iq_i+1=1$ for every $i=1,\ldots,n$.
To prove the second part of the claim, we set $p = \varphi(P), q = \varphi(Q)$ and $r = \varphi(R)$.
\end{proof}

The next polynomial we define is fundamental for our algebraic representation. This allows us to precisely construct our varieties and to study them.

\begin{Lemma}\label{lem:setpoly}
    Let $Q \subseteq \mathcal{P}$. The polynomial $\xi_Q \in \mathbb{B}(\bar{Y})$, defined as
        \begin{equation}
            \xi_Q(Y_1,\ldots,Y_n) = \prod_{i=1}^n (1 + Y_i + \varphi(Q)_i),
        \end{equation}
        has value $0$ at every point of $\mathbb{B} ^{n}$ except for $\varphi(Q)$ where it assumes value $1$. We call this the \emph{characteristic polynomial} of $Q$.
\end{Lemma}
\begin{proof}
    The evaluation $\xi_Q(\varphi(Q)) = \prod 1 = 1$. Let now $p \in \mathbb{B} ^{n}$ be such that $p \neq \varphi(Q)$; it exists $j \in \{ 1,\ldots,n \}$ for which $\varphi(Q)_j \neq p_j$. Thus the factor $1 - (p_j - \varphi(Q)_j)$ vanishes implying $\xi_Q(p) = 0$.
\end{proof}

In other words, the characteristic polynomial of a set $Q$ is a polynomial that has as zeros vectors of the form $v \in \mathbb{B}$ such that $\varphi^{-1}(v) \neq Q$. Equivalently we can say that $\xi_Q$ has $\varphi(Q)$ as unique non-zero.
In the next corollary, we show, given the characteristic polynomial $\xi_Q$ of a set $Q$, how we can obtain the characteristic polynomial of $Q^c$.

Let $\mathscr{R} = \{ \xi_S : S \subseteq \mathcal{P} \} \subseteq \mathbb{B}(\bar{Y})$ be the set given by all the characteristic polynomials of each subset $S \subseteq \mathcal{P}$.

\begin{corollary}
\label{cor:complem}
    Let $\xi_Q \in \mathscr{R}$, then
    \begin{equation}\label{eq:charcomp}
        \xi_{Q^c} = \frac{\xi_{\emptyset} \cdot \xi_{\mathcal{P}}}{\xi_{Q}}.
    \end{equation}
\end{corollary}
\begin{proof}
    Notice that $\xi_\emptyset = \prod_{i=1}^n (1 + Y_i)$ while $\xi_{\mathcal{P}} = Y_1\cdots Y_n$. Let $Q = \{P_{i_1},\ldots,P_{i_m}\}$, we have that 
    \begin{equation}
        \xi_Q = Y_{i_1}\cdots Y_{i_m} \cdot \prod_{j \not\in \{i_1,\ldots,i_m\}}(1+Y_j).
    \end{equation}
    Equation \ref{eq:charcomp} produces the polynomial
    \begin{equation}
        \left(\prod_{j \neq i_1,\ldots,i_m} Y_j \right) \cdot (1+Y_{i_1})\cdot \cdots \cdot(1+Y_{i_m}) = \xi_{Q^c}.
    \end{equation}
\end{proof}

\begin{remark}
    We have that $\xi_{Q^c} \in \mathbb{B}(\bar{X})$ since all of its terms are multilinear in the set of variables $\bar{Y}$.
\end{remark}

In Appendix~\ref{appendix:A} we give further results and constructions using characteristic polynomials. Moreover with the operations we define, the set of all the characteristic polynomials becomes a Boolean ring.

Let $\mathcal{A} \subseteq 2^\mathcal{P}$ and $\bar{Z} = \{Z_1,\ldots,Z_n\}$ be a set of variables. We denote with $\xi_{\mathcal{A}_{\bar{Z}}}$ the characteristic polynomial of $\mathcal{A}$ in the $\bar{Z}$ variables, defined as
\begin{equation}
    \xi_{\mathcal{A}_{\bar{Z}}} = \prod_{A \in \mathcal{A}} (\xi_A(\bar{Z}) + 1).
\end{equation}

\begin{remark}
\label{rem:varphi}
    It is easy to see that $\mathscr{V}(\langle \xi_A(\bar{Z}) + 1 \rangle) = \varphi(A)$, therefore by the product of ideals we obtain
    \begin{equation}
        \mathscr{V}( \langle \xi_{\mathcal{A}_{\bar{Z}}} \rangle) = \bigcup_{A \in \mathcal{A}}\mathscr{V}(\langle \xi_{A_{\bar{Z}}} + 1 \rangle) = \{ \varphi(A) : A \in \mathcal{A}\} = \varphi(\mathcal{A}).
    \end{equation}
\end{remark}

\section{Algebraic model for quorum systems}
Let $\mathcal{P}$ be a set of $n \in \mathbb{N}$ processes in a system. 
\label{sec:4}
\begin{definition}[Fail-prone system]
     A \emph{fail-prone system} $\mathcal{F} \subseteq 2^{\mathcal{P}}$ is a collection of subsets of $\mathcal{P}$, none of which is contained in another, such that some $F \in \mathcal{F}$ with $F \subseteq \mathcal{P}$ is called a \emph{fail-prone set} and contains all processes that may at most fail together in some execution.
\end{definition}

Henceforth, the notation $\mathcal{A}^*$ for a system $\mathcal{A} \subseteq 2^{\mathcal{P}}$ denotes the collection of all subsets of the sets in $\mathcal{A}$, that is, $\mathcal{A}^* =\{A'|A' \subseteq A,A \in \mathcal{A}\}$.
\subsection{Classical quorum system}

Classical quorum systems are applicable in the context of crash failures, i.e. when processes in the system can only stop executing steps \cite{DBLP:conf/sosp/Gifford79, Naor}.

\begin{definition}[Classical quorum system]\label{def:classquor}
A (classical) quorum system for a fail-prone system $\mathcal{F}$ is a collection of sets of processes $\mathcal{Q} \subseteq 2^\mathcal{P}$, where each $Q \in \mathcal{Q}$ is called a quorum, such that:
\begin{description}
    \item[Consistency] The intersection of any two quorums is non empty, i.e., 
    \begin{equation}
    \forall Q_1,Q_2 \in \mathcal{Q} : Q_1 \cap Q_2 \neq \emptyset.
\end{equation}
\item[Availability] For any set of processes that may fail together, there exists a disjoint quorum in $\mathcal{Q}$, i.e.,
    \begin{equation}
        \forall F \in \mathcal{F} :\exists~Q\in \mathcal{Q} : F\cap Q = \emptyset.
    \end{equation}
\end{description}
\end{definition}

Our first result expresses the consistency property of a quorum system using the tools we developed in previous sections.

\begin{theorem} 
\label{th:consclass}
Let $\mathcal{Q}$ be a quorum system and $\mathcal{F}$ be a fail-prone system. Consider the ideal $I = \langle \xi_{\mathcal{Q}_{\bar{X}}},\xi_{\mathcal{Q}_{\bar{Y}}}, \sigma \rangle \subseteq \mathbb{B}(\bar{X},\bar{Y})$ and let $\mathcal{G}$ be a Gr\"obner basis for $I$. Then, $\mathcal{Q}$ fulfills consistency with respect to $\mathcal{F}$ if
\begin{equation}
    |SM(\mathcal{G})| = |\mathcal{Q}|^2.
\end{equation}
\end{theorem}

\begin{proof}
    Recall that every ideal $J \subseteq \mathbb{B}(\bar{X},\bar{Y})$ is $0$-dimensional, therefore so is $I$. By Theorem~\ref{th:standard} we obtain that $|SM(\mathcal{G})| = |\mathscr{V}(I)|$. We analyze now the set $\mathscr{V}(I)$ and we show that $\mathscr{V}(I) \subseteq \varphi(\mathcal{Q})\times\varphi(\mathcal{Q})$.
    First of all we have that, since $I \subseteq \mathbb{B}(\bar{X},\bar{Y}) = \mathbb{F}_2[\bar{X},\bar{Y}]/\langle X_1^2+X_1,\ldots,X_n^2+X_n,Y_1^2+Y_1,\ldots,Y_n^2+Y_n \rangle$, then $\mathscr{V}(I) \subseteq \mathbb{B}^n \times \mathbb{B}^n$.\\
    From Remark~\ref{rem:varphi} we have that $\mathscr{V}(\langle \xi_{\mathcal{Q}_{\bar{X}}} \rangle) = \varphi(\mathcal{Q})\times \mathbb{B}^n$ and $\mathscr{V}(\langle \xi_{\mathcal{Q}_{\bar{Y}}} \rangle) = \mathbb{B}^n \times \varphi(\mathcal{Q})$. We take advantage of the sum of ideals rule to compute
    \begin{equation}
    \begin{split}
        \mathscr{V}(\langle \xi_{\mathcal{Q}_{\bar{X}}}, \xi_{\mathcal{Q}_{\bar{Y}}} \rangle) &= 
         \mathscr{V}(\langle \xi_{\mathcal{Q}_{\bar{X}}}\rangle + \langle \xi_{\mathcal{Q}_{\bar{Y}}} \rangle)\\
         &= \mathscr{V}(\langle \xi_{\mathcal{Q}_{\bar{X}}} \rangle) \cap \mathscr{V}(\langle \xi_{\mathcal{Q}_{\bar{Y}}} \rangle) = \\
         & = (\varphi(\mathcal{Q})\times \mathbb{B}^n) \cap (\mathbb{B}^n \times \varphi(\mathcal{Q}))\\
         & = \varphi(\mathcal{Q})\times\varphi(\mathcal{Q}).
    \end{split}
    \end{equation}
    Since $\langle \xi_{\mathcal{Q}_{\bar{X}}}, \xi_{\mathcal{Q}_{\bar{Y}}} \rangle \subseteq I$ then $\mathscr{V}(I) \subseteq \varphi(\mathcal{Q})\times\varphi(\mathcal{Q})$ as we claimed. Now, from a variety point of view, adding $\sigma$ to $\langle \xi_{\mathcal{Q}_{\bar{X}}}, \xi_{\mathcal{Q}_{\bar{Y}}} \rangle$, means to filter those vectors $v \in \varphi(\mathcal{Q})\times\varphi(\mathcal{Q})$ that satisfy
    \begin{equation}
        Supp(\pi_{\bar{X}}(v))\cap Supp(\pi_{\bar{Y}}(v)) \neq \emptyset
    \end{equation}
    as in Lemma~\ref{lem:supportint}.
    Through $\varphi^{-1}$ the two projections represent two quorums which intersect. Thus, consistency holds if $\mathscr{V}(I) = \varphi(\mathcal{Q})\times\varphi(\mathcal{Q})$. But then
    \begin{equation}
        |\mathscr{V}(I)| = | \varphi(\mathcal{Q})\times\varphi(\mathcal{Q})| = |\mathcal{Q}|^2.
    \end{equation}
    This proves the theorem.
\end{proof}

In other words, Theorem~\ref{th:consclass} shows that if common zeros of $\xi_{\mathcal{Q}_{\bar{X}}},\xi_{\mathcal{Q}_{\bar{Y}}}$ and $\sigma$ cover $\varphi(\mathcal{Q}) \times \varphi(\mathcal{Q})$, then the consistency property holds. This follows from the fact that covering $\varphi(\mathcal{Q}) \times \varphi(\mathcal{Q})$ means that every quorum in $\mathcal{Q}$ has a non empty pair-wise intersection. 

We can state the complementary result by constructing the ideal $I$ of Theorem~\ref{th:consclass} by substituting $\sigma$ with $\sigma + 1$. This allows us to skip the computation of the integer $|SM(\mathcal{G})|$.

\begin{corollary}
\label{cor:conclassrefined}
Let $\mathcal{Q}$ and $\mathcal{F}$ as in Theorem~\ref{th:consclass}. Consider the ideal $I = \langle \xi_{\mathcal{Q}_{\bar{X}}},\xi_{\mathcal{Q}_{\bar{Y}}}, \sigma + 1 \rangle$ and let $\mathcal{G}$ be a Gr\"obner basis for $I$. Then, $\mathcal{Q}$ fulfills consistency with respect to $\mathcal{F}$ if $\mathcal{G} = \{1\}$.
\end{corollary}

 Corollary \ref{cor:conclassrefined} relies only on the computation of a Gr\"obner basis of $I$ in order to enforce consistency of $\mathcal{Q}$ with respect to $\mathcal{F}$.

\begin{theorem}\label{th:availclass}
Let $\mathcal{Q}$ and $\mathcal{F}$ be a quorum system and a fail-prone system. Consider the ideal $I = \langle \xi_{\mathcal{F}_{\bar{X}}}, \xi_{ \mathcal{Q}_{\bar{Y}}}, \sigma +1 \rangle \subseteq \mathbb{B}(\bar{X},\bar{Y})$ and $\mathcal{G}$ be a Gr\"obner basis for $I$. Then $\mathcal{Q}$ fulfills availability with respect to $\mathcal{F}$ if
\begin{equation}\label{eq:dissavail}
        |SM(\mathcal{G}\cap \mathbb{B}(\bar{X}))| = |\mathcal{F}|.
    \end{equation}
\end{theorem}

\begin{proof}

With the same reasoning as in Theorem~\ref{th:consclass} we can prove that $\mathscr{V}(\langle \xi_{\mathcal{F}_{\bar{X}}}, \xi_{ \mathcal{Q}_{\bar{Y}}} \rangle) = \varphi(\mathcal{F})\times \varphi(\mathcal{Q})$ and moreover adding $\sigma + 1$ means to filter on vectors $v \in \mathscr{V}(I)$ such that 
    \begin{equation}
        Supp(\pi_{\bar{X}}(v))\cap Supp(\pi_{\bar{Y}}(v)) = \emptyset.
    \end{equation}
     Through $\varphi^{-1}$ the two projections represent a fail-prone set and a quorum that do not intersect. For the availability to hold we need $\pi_{\bar{X}}(\mathscr{V}(I)) = \varphi(\mathcal{F})$. This is enough because we know from Theorem~\ref{th:projvariety} that $\pi_{\bar{X}}(\mathscr{V}(I)) = \mathscr{V}(I \cap \mathbb{B}(\bar{X}))$ and from Theorem~\ref{th:extension} that, for every $p \in \mathscr{V}(I \cap \mathbb{B}(\bar{X}))$ there exists $q \in \mathbb{B}^n$ such that $(p,q) \in \mathscr{V}(I)$. In other words, through $\varphi^{-1}$, $q$ is the representation of a quorum not intersecting the fail-prone represented by $p$. As \protect{\cite[Theorem~2.3.4]{loustaunau1994introduction}} says, $\mathcal{G}\cap \mathbb{B}(\bar{X})$ is a Gr\"obner basis for $I \cap \mathbb{B}(\bar{X})$ meaning that we can apply Theorem~\ref{th:standard} to state
     \begin{equation}
         |SM(\mathcal{G}\cap \mathbb{B}(\bar{X}))| = |\pi_{\bar{X}}(\mathscr{V}(I))| = |\varphi(\mathcal{F})|
     \end{equation}
     proving our thesis.
\end{proof}

Theorems \ref{th:consclass} and \ref{th:availclass} prove that there is a relation between the consistency and availability properties and the size of the standard monomials of the related Gr\"obner basis. Results in this section are proved without ever checking the varieties of the constructed ideals, instead the Gr\"obner bases computation is required. We can therefore state whether a set of sets $\mathcal{Q}$ is a quorum system, with respect to a second set of sets $\mathcal{F}$, just by inspecting the leading monomials of the related Gr\"obner bases.
In the next sections we use the same arguments in order to extend our result to other quorum systems.

\subsection{Dissemination quorum system}
Malkhi and Reiter \cite{DBLP:journals/dc/MalkhiR98} introduced a generalization of classical quorum systems called Byzantine quorum systems. These are useful in systems that may be subject to arbitrary (or Byzantine) failures, i.e., if a process may deviate in any conceivable way from the algorithm assigned to it \cite{DBLP:books/daglib/0025983}.
They presented two kinds of quorum systems, namely \emph{dissemination quorum systems} and \emph{masking quorum systems}. The former aims at storing self-verifying (or authenticated) data in a replicated system, whereas the latter one has the goal of storing unauthenticated data~\cite{DBLP:journals/eatcs/Vukolic10}. They have found many more applications in distributed protocols.

\begin{definition}[Dissemination quorum system]\label{def:dissemin}
A (Byzantine) dissemination quorum system for a fail-prone system $\mathcal{F}$ is a collection of sets of processes $\mathcal{Q} \subseteq 2^\mathcal{P}$, where each $Q \in \mathcal{Q}$ is called a quorum, such that the following properties hold:
\begin{description}
    \item[Consistency] The intersection of any two quorums contains at least one process that is not fail-prone, i.e., 
    \begin{equation}
    \forall Q_1,Q_2 \in \mathcal{Q}, \forall F \in \mathcal{F} : Q_1 \cap Q_2 \not\subseteq F.
\end{equation}
\item[Availability] For any set of processes that may fail together, there exists a disjoint quorum in $\mathcal{Q}$, i.e.,
    \begin{equation}
        \forall F \in \mathcal{F} :\exists~Q\in \mathcal{Q} : F\cap Q = \emptyset.
    \end{equation}
\end{description}
\end{definition}

Exploiting the properties of $\gamma$ we define the Boolean polynomial $\gamma(\varphi(F), \bar{T})$ for every $F \in \mathcal{F}$ such that its zero locus are the points representing the powerset $2^F$. We can thus express the consistency property in (\ref{def:dissemin}) like in the previous section, i.e., by defining an ideal and checking its variety's properties. 

\begin{Lemma}\label{lem:lambda}
    The zero locus of the polynomial $\lambda \in \mathbb{B}(\bar{T})$, defined as
    \begin{equation}
        \lambda = \prod_{F\in \mathcal{F}}\gamma(\varphi(F),\bar{T}),
    \end{equation}
    is the set of points of $\mathbb{B} ^{n}$ representing the elements of  $\mathcal{F}^*$.
\end{Lemma}
\begin{proof}
    From Corollary~\ref{lem:support}, we have that $F' \subseteq F$ if and only if
    \begin{equation}
     \gamma(\varphi(F),\varphi(F')) = 0 
    \end{equation}
    where $\varphi(F')$ is a point of $\mathbb{B} ^{n}$ representing a subset of $F$. It follows that the zero locus of $\lambda$ is the set of points of $\mathbb{B} ^{n}$ representing the elements of $\mathcal{F}^*$.
\end{proof}

The consistency property of dissemination quorum systems states that the pair-wise intersections of quorums in $\mathcal{Q}$ are not contained in any fail-prone set. We consider the two characteristic polynomials $\xi_{\mathcal{Q}_{\bar{X}}}$ and $\xi_{\mathcal{Q}_{\bar{Y}}}$ along with $\lambda$ as in Lemma~\ref{lem:lambda} and $\delta$ as in (\ref{eq:delta}).
    
\begin{theorem}
\label{th:consdissem}
Let $\mathcal{Q}$ and $\mathcal{F}$ be a quorum system and a fail-prone system. Let $I$ be the ideal $I = \langle \xi_{\mathcal{Q}_{\bar{X}}}, \xi_{\mathcal{Q}_{\bar{Y}},}, \lambda, \delta \rangle \subseteq \mathbb{B}(\bar{X},\bar{Y}, \bar{T})$ and $\mathcal{G}$ be a Gr\"obner basis for $I$. We say that $\mathcal{Q}$ fulfills consistency with respect to $\mathcal{F}$ if
     \begin{equation}
         |SM(\mathcal{G})| = |\mathcal{Q}|^2 \cdot |\mathcal{F}^*|.
     \end{equation}
\end{theorem}
\begin{proof}
Since $I \subseteq \mathbb{B}(\bar{X},\bar{Y}, \bar{T}) = \mathbb{F}_2[\bar{X},\bar{Y}, \bar{T}]/\langle X_1^2+X_1,\ldots,X_n^2+X_n,Y_1^2+Y_1,\ldots,Y_n^2+Y_n,T_1^2+T_1,\ldots,T_n^2+T_n \rangle$, then $\mathscr{V}(I) \subseteq \mathbb{B}^n \times \mathbb{B}^n \times \mathbb{B}^n$.\\
With the same reasoning as in Theorem~\ref{th:consclass}, noticing that $\mathscr{V}(\langle \lambda \rangle) = \varphi(\mathcal{F}^*) \times \mathbb{B}^n$, it is possible to prove that $\mathscr{V}(I) \subseteq \varphi(\mathcal{Q}) \times \varphi(\mathcal{Q}) \times \varphi(\mathcal{F}^*)$. Then, adding $\delta$ to $\langle \xi_{\mathcal{Q}_{\bar{X}}}, \xi_{\mathcal{Q}_{\bar{Y}},}, \lambda \rangle$, means to filter those vectors $v \in \varphi(\mathcal{Q}) \times \varphi(\mathcal{Q}) \times \varphi(\mathcal{F}^*)$ that satisfy 
\begin{equation}
    (\mathrm{Supp}(p) \cap \mathrm{Supp}(q)) \not\subseteq \mathrm{Supp}(r)
\end{equation}
as in Lemma~\ref{lem:supportdissemin}. Consistency property then holds if $\mathscr{V}(I) = \varphi(\mathcal{Q}) \times \varphi(\mathcal{Q}) \times \varphi(\mathcal{F}^*).$ It follows that $|SM(\mathcal{G})| = |\mathscr{V}(I)| = |\mathcal{Q}|^2 \cdot |\mathcal{F}^*|$ Theorem follows.
 
\end{proof}

In other words, Theorem~\ref{th:consdissem} shows that there is a relationship between common zeros of $\xi_{\mathcal{Q}_{\bar{X}}},\xi_{\mathcal{Q}_{\bar{Y}}}, \lambda$ and $\delta$ and consistency property. Asking $\mathscr{V}(\langle \xi_{\mathcal{Q}_{\bar{X}}}, \xi_{\mathcal{Q}_{\bar{Y}}}, \lambda, \delta \rangle)$ to cover $\varphi(\mathcal{Q})^2 \times \varphi(\mathcal{F}^*)$ means that every pair of quorums and every fail-prone are zeros of $\delta$ which represent the condition required for the consistency property of dissemination quorum systems. 

Malkhi and Reiter proved that dissemination quorum systems can only exist if not \emph{too many} processes fail \cite{DBLP:journals/dc/MalkhiR98}. Let us define the $Q^3$ condition~\cite{DBLP:journals/joc/HirtM00}. 

\begin{definition}[$Q^3$-condition] A fail prone system $\mathcal{F}$ satisfies the $Q^3$-condition, abbreviated as $Q^3(\mathcal{F})$, whenever it holds,
    \begin{equation}
        \forall F_1,F_2,F_3 \in \mathcal{F} : \mathcal{P} \not \subseteq F_1 \cup F_2 \cup F_3.
    \end{equation}
\end{definition}

Loosely speaking, the $Q^3$ conditions ensures that there is no combination of three fail-prone sets that can cover the entire set of players.

\begin{Lemma}[\protect{\cite[Theorem~5.4]{DBLP:journals/dc/MalkhiR98}}]
\label{lem:canon}
  Let $\mathcal{F}$ be a fail-prone system. A dissemination quorum system for $\mathcal{F}$ exists
  if and only if~$Q^3(\mathcal{F})$.
\end{Lemma}

We rephrase the $Q^3$-condition in algebraic terms. Define the polynomial $\omega \in \mathbb{B}(\bar{X},\bar{Y},\bar{T})$, as
\begin{equation}
    \omega(\bar{X},\bar{Y},\bar{T}) = \prod_{i=1}^n (X_iY_iT_i + X_iY_i + X_iT_i + Y_iT_i + X_i + Y_i + T_i).
\end{equation}

\begin{Lemma}\label{lem:fullunion}
    Given $a,b,c \in \mathbb{B} ^{n}$ we have that $Supp(a) \cup Supp(b) \cup Supp(c) = \{1,\ldots,n \}$ if and only if $\omega(a,b,c) = 1$.
\end{Lemma}
\begin{proof}
Assume first $Supp(a) \cup Supp(b) \cup Supp(c) = \{1,\ldots,n \}$ , this means that for every $i = 1,\ldots,n$ at least one of $a_i,b_i$ and $c_i$ is $1$ thus the factor $X_iY_iT_i + X_iY_i + X_iT_i + Y_iT_i + X_i + Y_i + T_i$ evaluates to $1$. This implies $\omega(a,b,c) = 0$.
We can prove the other way around with the reverse argumentation.
\end{proof}

Next theorem gives an algebraic way, following the same idea of previous section, to check $Q^3$ condition inspecting properties of a specific ideal.

\begin{theorem}
Given a fail prone system $\mathcal{F}$, consider the ideal $I = \langle \xi_{\mathcal{F}_{\bar{X}}},\xi_{\mathcal{F}_{\bar{Y}}},\xi_{\mathcal{F}_{\bar{T}}}, \omega \rangle \subseteq \mathbb{B}[\bar{X},\bar{Y},\bar{T}]$. Let $\mathcal{G}$ be a Gr\"obner basis for $I$. Then $\mathcal{F}$ satisfies $Q^3(\mathcal{F})$ if 
\begin{equation}
    |SM(\mathcal{G})| = |\varphi(\mathcal{F})|^3.
\end{equation}
\end{theorem}
\begin{proof}
    Apply the same arguments as in Theorem~\ref{th:consclass}.
\end{proof}

\begin{remark}
Observe that, under the threshold failure model, we can express a quorum system as 
$\mathcal{Q} = \{\xi \in \mathscr{R} | \deg(TM(\xi)) \ge \frac{n+f+1}{2}\}$ where $f$ is the number of processes that may fail together. Furthermore, a fail-prone system is $ \mathcal{F} = \{\xi \in \mathscr{R} | \deg(TM(\xi)) \le f\}$.
\end{remark}

\subsection{Masking quorum system}

\begin{definition}[Masking quorum system]\label{def:masking}
A (Byzantine) masking quorum system for a fail-prone system $\mathcal{F}$ is a collection of sets of processes $\mathcal{Q} \subseteq 2^\mathcal{P}$, where each $Q \in \mathcal{Q}$ is called a quorum, such that the following properties hold:
\begin{description}
    \item[Consistency] The intersection of any two quorums contains at least one process that is not fail-prone even when removing from the intersection another fail-prone set, i.e., 
    \begin{equation}
    \forall Q_1,Q_2 \in \mathcal{Q}, \forall F_1, F_2 \in \mathcal{F} : (Q_1 \cap Q_2) \setminus F_1 \not\subseteq F_2.
\end{equation}
\item[Availability] For any set of processes that may fail together, there exists a disjoint quorum in $\mathcal{Q}$, i.e.,
    \begin{equation}
        \forall F \in \mathcal{F} :\exists Q\in \mathcal{Q} : F\cap Q = \emptyset.
    \end{equation}
\end{description}
\end{definition}

We formulate the consistency property in masking quorum system algebraically. Assuming $\mathcal{F} = \{F_1,\ldots,F_m\}$, in the next theorem we denote $\mathcal{F}^c$ the set $\{F_1^c,\ldots,F_m^c\}$. 
\begin{theorem}\label{th:consmask}
Let $\mathcal{Q}$ and $\mathcal{F}$ be a quorum system and a fail-prone system. Let $I$ be the ideal $I = \langle \xi_{\mathcal{Q}_{\bar{X}}}, \xi_{\mathcal{Q}_{\bar{Y}}}, \xi_{\mathcal{F}^c_{\bar{Z}}}, \lambda, \delta' \rangle \subseteq \mathbb{B}(\bar{X},\bar{Y}, \bar{Z}, \bar{T})$, where $\delta'(\bar{X},\bar{Y},\bar{Z},\bar{T}) =\prod_{i = 1}^n(T_iX_iY_iZ_i +X_iY_iZ_i + 1)$, and $\mathcal{G}$ be a Gr\"obner basis for $I$. We say that $\mathcal{Q}$ fulfills consistency with respect to $\mathcal{F}$ if
 \begin{equation}
     |SM(\mathcal{G})| = |\mathcal{Q}|^2\cdot |\mathcal{F}|\cdot |\mathcal{F}^*|.
 \end{equation}
\end{theorem}
\begin{proof}
    Apply the same reasoning as in Theorem~\ref{th:consdissem}.
\end{proof}

Theorem~\ref{th:consmask} follows the same approach as Theorem~\ref{th:consclass} and Theorem~\ref{th:consdissem} by showing the relationship between elements of $\mathscr{V}(\langle \xi_{\mathcal{Q}_{\bar{X}}}, \xi_{\mathcal{Q}_{\bar{Y}}}, \xi_{\mathcal{F}^c_{\bar{Z}}},\lambda, \delta' \rangle)$ and the consistency property. Notice that we express consistency property by using the equivalence $(A \cap B) \setminus C = (A \cap B) \cap C^c$, with $A,B$ and $C$ sets.

\begin{remark}
Malkhi and Reiter \cite{DBLP:journals/dc/MalkhiR98} proved a similar condition as $Q^3$ for masking quorum systems called $Q^4$. This is essentially the same except for quantification over fail-prone sets. In this case we say that $\mathcal{F}$ satisfies $Q^4(\mathcal{F})$ whenever it holds
    \begin{equation}
        \forall F_1,F_2,F_3,F_4 \in \mathcal{F} : \mathcal{P} \not \subseteq F_1 \cup F_2 \cup F_3 \cup F_4.
    \end{equation}
We omit its algebraic construction as it is similar as the one presented for dissemination quorum systems.
\end{remark}

\section{Algorithms}
\label{sec:algo}

The present section introduces the basic algorithm for Gr\"obner basis computation introduced by Buchberger. Such algorithm and more sophisticated ones are implemented in many symbolic computer algebra systems like PolyBoRi \cite{DBLP:journals/jsc/BrickensteinD09}, BooleanBG \cite{DBLP:journals/corr/abs-1010-2669}, Maculay2 \cite{Macaulay2} and Magma \cite{MR1484478}.
Afterwards we give an algorithm that makes use of Corollary~\ref{cor:conclassrefined} to characterize consistency in classical quorum systems. For the sake of completeness, we will first give some definitions that we will need to introduce the algorithms. The definitions we mention can be found in the standard literature \cite{DBLP:conf/carla/CamposTM16}.

\begin{definition} Let $f_1, f_2 \in \mathbb{B}(\bar{X})$. We say that $f_1$ is reducible by $f_2$ if
$LM(f_2) | LM(f_1)$. The reduction of $f_1$ by $f_2$ is defined as $\textsl{red}(f_1, f_2) := f_1 - \frac{LM(f_1)}{LM(f_2)}$.
\end{definition}
\begin{definition} Let $f_1 \in \mathbb{B}(\bar{X})$ and $S \subseteq \mathbb{B}(\bar{X})$. The reduction
of $f_1$ by $S$ is defined as $\textsl{red}(f_1,S) := \textsl{red}(\textsl{red}(f_1, S_i),S \setminus \{S_i\})$ if it is possible to
choose some $S_i \in S$ as a valid reductor and $f_1$ otherwise.
\end{definition}
\begin{definition} Let $f_1, f_2 \in \mathbb{B}(\bar{X})$. The s-polynomial of $f_1$ and $f_2$ is
defined as $\textsl{sp}(f_1, f_2) := \frac{\lambda}{LM(f_1)}f_1 + \frac{\lambda}{LM(f_2)}f_2$ where $\lambda = LCM(LM(f_1), LM(f_2))$.
\end{definition}
\begin{definition} Let $\mathcal{G} \subseteq \mathbb{B}(\bar{X})$ be a basis of $I$. $\mathcal{G}$ is a Gr\"obner basis
of $I$ if $\forall g_i, g_j \in \mathcal{G}, \textsl{red}(\textsl{sp}(g_i, g_j),\mathcal{G}) = 0$.
\end{definition}

In his Ph.D. thesis \cite{Buc65}, Buchberger designed also two criteria to characterize when a s-polynomial reduces to zero, since these reductions do not give contributions to the computation of a Gr\"obner basis. We will call the two criteria \emph{coprime criterion} and \emph{chain criterion}, respectively.

\begin{theorem}[Coprime criterion]
Let $f_1, f_2 \in \mathcal{G}$ and $\mathcal{G} \subseteq \mathbb{B}(\bar{X})$. The polynomial $\textsl{sp}(p, q)$
will reduce to zero if $LM(f_1)$ and $LM(f_2)$ are coprime.
\end{theorem}
\begin{theorem}[Chain criterion]
Let $f_1, f_2 \in \mathcal{G}$ and $\mathcal{G} \subseteq \mathbb{B}(\bar{X})$. The polynomial $\textsl{sp}(p, q)$ will reduce to zero if $\exists g \in \mathcal{G} : LM(g) | LM(\textsl{sp}(f_1, f_2))$ and $\textsl{red}(\textsl{sp}(f_1, g),\mathcal{G}) =
\textsl{red}(\textsl{sp}(f_2, g),\mathcal{G}) = 0$.
\end{theorem}

We give Buchberger's algorithm which takes as input a basis $\mathcal{G}$ for an ideal $I \subseteq \mathbb{B}(\bar{X})$ and outputs a Gr\"obner basis for $I$. We always assume the usage of the ordering \textsc{lex}.

\begin{algorithm}
\caption{}
    \begin{algorithmic}[1]
    \Function{Buchberger}{$\mathcal{G} \subseteq \mathbb{B}(\bar{X})$}
        \State $S \leftarrow \{(g_i, g_j) : g_i, g_j \in \mathcal{G}, j>i\}$
        \While{$S \neq \emptyset$} 
            \State $s \leftarrow \textsl{select}(S)$
            \State $S \leftarrow S \setminus \{s\}$
            \If{$\lnot \textsl{coprime}(s_1, s_2) \land \lnot \textsl{chain}(s_1, s_2, \mathcal{G})$} 
                \State $r \leftarrow \textsl{red}(\textsl{sp}(s_1, s_2), \mathcal{G})$
                \If{$r \neq 0$} 
                    \State $S \leftarrow S \cup \{ (r,g) : g \in \mathcal{G} \}$
                    \State $\mathcal{G} \leftarrow \mathcal{G} \cup \{r\}$
                \EndIf
            \EndIf
        \EndWhile
        \State \Return $\mathcal{G}$
    \EndFunction
    \end{algorithmic}
        \label{alg:buchberger}
    \end{algorithm}

An efficient implementation of the chain criterion was introduced by Gabauer and M\"oller~\cite{DBLP:journals/jsc/GebauerM88}, and recently improved by Campos \cite{DBLP:conf/carla/CamposTM16}. Moreover, a detailed analysis along with benchmarks, of existing algorithms for computing Boolean Gr\"obner basis, is reported in \protect{\cite[Section~5]{DBLP:conf/carla/CamposTM16}}.
The consistency property can now be tested with an algorithm that takes as input the set $\mathcal{Q}$ and outputs an element of $\{\textsc{true}, \textsc{false}\}$.
Notice that we can also consider the computations of all the characteristic polynomials as a preprocessing step, therefore the complexity of the algorithm only relies on the Gr\"obner basis computation in Algorithm~\ref{alg:buchberger}.

\begin{algorithm}
\caption{}
    \begin{algorithmic}[1]
    \Function{Consistency}{$\mathcal{Q}$}
        \State $\xi_{\bar{X}}, \xi_{\bar{Y}} \leftarrow 1$
        \State $\sigma \leftarrow \prod_{i=1}^n (X_iY_i +1)$
        \ForAll{$Q \in \mathcal{Q}$}
            \State $\xi_{\bar{X}} \leftarrow \xi_{\bar{X}} \cdot \prod_{i=1}^n (1 + X_i + \varphi(Q)_i)$
            \State $\xi_{\bar{Y}} \leftarrow \xi_{\bar{Y}} \cdot \prod_{i=1}^n (1 + Y_i + \varphi(Q)_i)$
        \EndFor
        \State $\mathcal{G} \leftarrow \textsc{Buchberger}(\{ \xi_{\bar{X}}, \xi_{\bar{Y}}, \sigma + 1 \})$
        \State \Return $\mathcal{G} \stackrel{?}{=}\{ 1 \}$
    \EndFunction
    
    \end{algorithmic}
    \label{alg:consistency}
    \end{algorithm}

Next we sketch an algorithm that uses techniques described in Theorem \ref{th:availclass}. It tests availability of a quorum system $\mathcal{Q}$ with respect to a fail-prone system $\mathcal{F}$. 

\begin{algorithm}
\caption{}
    \begin{algorithmic}[1]
    
    \Function{Availability}{$\mathcal{Q}, \mathcal{F}$}
        \State $\xi_{\mathcal{F}}, \xi_{\mathcal{Q}} \leftarrow 1$
        \State $\sigma \leftarrow \prod_{i=1}^n (X_iY_i +1)$
        \State $\mathcal{G}' \leftarrow \emptyset$
        \ForAll{$F \in \mathcal{F}$}
            \State $\xi_{\mathcal{F}} \leftarrow \xi_{\mathcal{F}} \cdot \prod_{i=1}^n (1 + X_i + \varphi(F)_i)$
        \EndFor
        \ForAll{$Q \in \mathcal{Q}$}
            \State $\xi_{\mathcal{Q}} \leftarrow \xi_{\mathcal{Q}} \cdot \prod_{i=1}^n (1 + Y_i + \varphi(Q)_i)$
        \EndFor
        \State $\mathcal{G} \leftarrow \textsc{Buchberger}(\{ \xi_{\mathcal{F}}, \xi_{\mathcal{Q}}, \sigma + 1 \})$
         \ForAll{$g \in \mathcal{G}$}
            \If{$g \in \mathbb{B}[X_1,\ldots,X_n]$} 
            \State $\mathcal{G}' \leftarrow \mathcal{G}' \cup \{g\}$
        \EndIf
        \EndFor
        \State $\textit{Standard} = SM(\mathcal{G}')$
        \State \Return $|\textit{Standard}| \stackrel{?}{=} |\mathcal{F}|$
    \EndFunction
    
    \end{algorithmic}
    \label{alg:availability}
\end{algorithm}

Further algorithms implementing theorems in Section~4 can be devised following the structure of Algorithm~\ref{alg:consistency} and Algorithm~\ref{alg:availability}.
A possible method for computing the $SM(\mathcal{G})$ function can be implemented evaluating the numerator of the Hilbert series of the Stanley-Reisner ring \cite{miller2004combinatorial}. They use a combinatorial argument on simplicial topology representing monomial ideals to study their structure.
We want to stress that the time complexity of the proposed algorithm entirely depend on the complexity of the Buchberger algorithm and on the $SM(\mathcal{G})$ algorithm.

\section{Conclusions and future work}
\label{sec:conc}

In this work we took advantage of well-known algebraic techniques in order to express properties of different quorum systems. We proved that given a custom set of sets, one can, in principle, test using Gr\"obner bases whether the set fulfills the requirements of a quorum system.
We leave it for future research to actually evaluate the complexity of our method and to explore potential optimizations, in terms of time and memory consumption. Furthermore, we strongly believe that refinements of our main results are possible, proving for instance that the conditions we give are not only necessary but also sufficient. Devising an actual algorithm to implement $SM(\mathcal{G})$ is a natural next step of this work.

Traditionally, trust assumption has been symmetric, in which all processes have to adhere on a global fail-prone structure. Damg{\aa}rd \emph{et al.}~\cite{DBLP:conf/asiacrypt/DamgardDFN07} introduced an \emph{asymmetric} trust assumption, in which every process is allowed to trust on a personal failing structure. Cachin and Tackmann~\cite{DBLP:conf/opodis/CachinT19} introduced asymmetric Byzantine quorum systems as a generalization of Byzantine quorums systems for asymmetric trust. An asymmetric fail-prone system $\mathbb{F}$ consists of an array of fail-prone systems, one for every process $p_i$ in the system. An \emph{asymmetric Byzantine quorum system} $\mathbb{Q}$ for $\mathbb{F}$ is an array of quorum systems, one for every process $P_i$ such that, in a similar way as in the symmetric case, the intersection of two quorums for any two processes contains at least one process for which both processes assume that it is not faulty \emph{and} for any process $P_i$ and any set of processes that may fail together according to $P_i$, there exists a disjoint quorum for $P_i$ in its quorum system.

Another approach to asymmetric trust was proposed by the Stellar blockchain. The Stellar consensus protocol~\cite{Mazieres2015TheSC} powers the Stellar Lumen (XLM) cryptocurrency and introduces federated Byzantine quorum systems (FBQS).  FBQS rely on the concept of a \emph{quorum slice}, which is a subset of the processes that can convince one particular process of agreement. According to the formalization of Stellar, a quorum as a non-empty set $Q \subset 2^{\mathcal{P}}$ that contains at least one quorum slice for each of its non-faulty members. 

An algebraic model of these two approaches appears interesting and feasible. The ultimate goal will be to formulate a comprehensive model of the symmetric and asymmetric quorum-system worlds without referring to set-system properties. We believe this will help finding new and different algorithms for implementing quorums in real-world distributed systems.

\section*{Acknowledgments}

The authors would like to express their great appreciation to Christian Cachin, Giorgia Azzurra Marson and Alessio Meneghetti for their valuable and constructive suggestions during the planning and development of this research work. 

This work has been funded by the Swiss National Science Foundation (SNSF)
under grant agreement Nr\@.~200021\_188443 (Advanced Consensus Protocols).

\appendix
\section{Operations on characteristic polynomials}
\label{appendix:A}

We present some results on characteristic polynomials. In particular, we show how to construct characteristic polynomials of intersection and union of sets, and lately on sets of shape $(A \cap B)\setminus C$. 

\begin{Lemma}\label{lem:trailing}
    Let $Q = \{P_{i_1},\ldots,P_{i_m}\} \in \mathcal{P}$, then
    \begin{equation}
        TM(\xi_{Q}) = Y_{i_1}\cdots Y_{i_m}
    \end{equation}
\end{Lemma}
\begin{proof}
The trailing monomial is obtained by the multiplication of $Y_{i_1}\cdots Y_{i_m}$ and the $1$s in $(1-Y_{j_1})\cdots (1-Y_{j_{n-m}})$. Thus $TM(\xi_{Q}) = Y_{i_1}\cdots Y_{i_m}$ 
\end{proof}

We start by constructing the characteristic polynomial of the intersection of two sets $Q$ and $R$.

\begin{proposition}
\label{prop:charint}
Let $Q,R \in \mathcal{P}$ and define

\begin{equation}
    \mu = \gcd(TM(\xi_{Q}), TM(\xi_{R}))
\end{equation}
and
\begin{equation}
    \nu = \frac{\xi_{Q}}{TM(\xi_{Q})}\cdot \frac{\xi_{R}}{TM(\xi_{R})}.
\end{equation}

Then,

\begin{equation}
    \xi_{Q \cap R} = \mu \cdot \nu
\end{equation}

\end{proposition}

\begin{proof}
   Assume $Q=\{P_i\}_{i \in I}$ and $R=\{P_j\}_{j \in J}$ and let $Q \cap R = \{P_k\}_{k \in K = I \cap J}$.
   We want to prove that 
   \begin{equation}\label{eq:munu}
       \xi_{Q \cap R} = \prod_{k \in K}Y_k \cdot \prod_{l \in \{1,\ldots,n\} \setminus K}(1 + Y_l) = \mu \cdot \nu
   \end{equation}
   From Lemma~\ref{lem:trailing} we have that $TM(Q) = \prod_{i \in I}Y_i$ and $TM(R) = \prod_{j \in J}Y_j$ and therefore $\mu = \gcd(TM(\xi_{Q}), TM(\xi_{R})) = \prod_{k \in K}Y_k$.
   Now
   \begin{equation}
       \frac{\xi_{Q}}{TM(\xi_{Q})} = \prod_{i \in \{1,\ldots,n\} \setminus I}(1 + Y_i)
   \end{equation}
   and the same holds for $\frac{\xi_{R}}{TM(\xi_{R})}$ over $J$.
   Notice that since $(\{1,\ldots,n\}\setminus I) \cup (\{1,\ldots,n\}\setminus J) = \{1,\ldots,n\}\setminus K$ and since we are working on the binary field, i.e. $(Y_i)^2=Y_i$, we can write
   \begin{equation}
        \nu = \frac{\xi_{Q}}{TM(\xi_{Q})}\cdot \frac{\xi_{R}}{TM(\xi_{R})} = \prod_{l \in \{1,\ldots,n\} \setminus K}(1 + Y_l)
   \end{equation}
   Thus the product $\mu \cdot \nu$ gives the equality in equation (\ref{eq:munu}).
\end{proof}

Furthermore, we present a construction for the union of $Q$ and $R$.

 \begin{proposition}
    Let $Q,R \in \mathcal{P}$ and define
    \begin{equation}
        \mu = TM(\xi_{Q}) \cdot TM(\xi_{R})
    \end{equation}
    and
    \begin{equation}
        \nu = \gcd(\frac{\xi_{Q}}{TM(\xi_{Q})}, \frac{\xi_{R}}{TM(\xi_{R})})
    \end{equation}
    Then,

    \begin{equation}
        \xi_{Q \cup R} = \mu \cdot \nu
    \end{equation}
\end{proposition}
\begin{proof}
    Apply the same argument as in Proposition~\ref{prop:charint} bearing in mind that if $A$ and $B$ are two monomials in $\mathbb{B}(Y_1,\ldots,Y_n)$ then $\lcm(A,B) = A\cdot B$.
\end{proof}

\begin{example}
\label{ex:int-un}
    Let $n = 6$, therefore $\mathcal{P} = \{P_1,\ldots,P_6\}$, and let $Q = \{P_2,P_3,P_4,P_6\}$ and $R = \{P_3,P_4,P_5\}$. Construct the characteristic polynomials as in Lemma~\ref{lem:setpoly}
    \begin{equation}
        \xi_{Q} = Y_1Y_2Y_3Y_4Y_5Y_6 + Y_1Y_2Y_3Y_4Y_6 + Y_2Y_3Y_4Y_5Y_6 + Y_2Y_3Y_4Y_6
    \end{equation}
    and
    \begin{equation}
        \begin{split}
            \xi_{R} =& Y_1Y_2Y_3Y_4Y_5Y_6 + Y_1Y_2Y_3Y_4Y_5 + Y_1Y_3Y_4Y_5Y_6 + Y_2Y_3Y_4Y_5Y_6 +\\  & Y_1Y_3Y_4Y_5 + Y_2Y_3Y_4Y_5 + Y_3Y_4Y_5Y_6 + Y_3Y_4Y_5
        \end{split}
    \end{equation}
    We obtain $TM(\xi_{Q}) = Y_2Y_3Y_4Y_6$ and $TM(\xi_{R})=Y_3Y_4Y_5$. Let us compute the characteristic polynomial of $Q \cup R$. First, compute $\mu$ and $\nu$ as the following.
    \begin{equation}
        \mu = \gcd(Y_2Y_3Y_4Y_6, Y_3Y_4Y_5) = Y_3Y_4
    \end{equation}
    and
    \begin{equation}
    \begin{split}
        \nu = &(Y_1Y_5 + Y_1 + Y_5 + 1)(Y_1Y_2Y_6 + Y_1Y_2 + Y_1Y_6 + Y_2Y_6 + Y_1 + Y_2 + Y_6 + 1)\\
        =& Y_1Y_2Y_5Y_6 + Y_1Y_2Y_5 + Y_1Y_2Y_6 + Y_1Y_5Y_6 + Y_1Y_2 + Y_1Y_5 + Y_1Y_6 + Y_2Y_5Y_6 + \\ 
        &   +Y_1 + Y_2Y_5 + Y_2Y_6 + Y_5Y_6 + Y_2 + Y_5 + Y_6 + 1
           \end{split}
    \end{equation}
    Finally, $(\mu\cdot \nu)(\bar{Y}) = 1$ only in $(0,0,1,1,0,0)$, i.e. $\varphi^{-1}(0,0,1,1,0,0) = \{P_3,P_4\} = Q \cap R$ meaning that $\mu\cdot \nu = \xi_{Q \cap R} = Y_3Y_4(1+Y_1)(1+Y_2)(1+Y_5)(1+Y_6)$.
    
    Now we compute the characteristic polynomial of $Q \cap R$. Again, let us compute $\mu$ and $\nu$ as the following.
        \begin{equation}
        \mu = Y_2Y_3^2Y_4^2Y_5Y_5=Y_2Y_3Y_4Y_5Y_6 
    \end{equation}
    and
    \begin{equation}
        \nu = \gcd(Y_1Y_5+Y_1+Y_5+1, Y_1Y_2Y_6+Y_1Y_2+Y_1Y_6+Y_2Y_6+Y_1+Y_2+Y_6+1)=(1+Y_1)
    \end{equation}
    Finally, $(\mu\cdot \nu)(\bar{Y}) = 1$ only in $(0,1,1,1,1,1)$. 
    
    It follows that $\varphi^{-1}(0,1,1,1,1,1) = \{P_2,P_3,P_4,P_5,P_6\} = Q \cup R$ meaning that $\mu\cdot \nu = \xi_{Q \cup R} = Y_1(1+Y_2)(1+Y_3)(1+Y_4)(1+Y_5)(1+Y_6)$.
    \qed
\end{example}

We show how is it possible to obtain the characteristic polynomial on \emph{more complex} sets as the following.

\begin{proposition}
\label{prop:charmasking}
Let $Q,R, F \subseteq \mathcal{P}$ and define

\begin{equation}
    \mu = \gcd\left(\frac{\xi_\mathcal{P}}{TM(\xi_F)},TM(\xi_{Q}), TM(\xi_{R})\right)
\end{equation}
and
\begin{equation}
    \nu = \frac{\xi_{Q}}{TM(\xi_{Q})}\cdot \frac{\xi_{R}}{TM(\xi_{R})}\cdot\frac{\xi_{F^c}}{TM(\xi_{F^c})}.
\end{equation}

Then,

\begin{equation}
    \xi_{(Q \cap R)\setminus F} = \xi_{(Q \cap R)\cap F^c} = \mu \cdot \nu
\end{equation}

\end{proposition}

\begin{proof}
    Apply the same argument as in Proposition~\ref{prop:charint} bearing in mind that $TM(\xi_{F^c})=\frac{\xi_{\mathcal{P}}}{TM(\xi_{F})}$, with $\xi_{F_c}$ as in Corollary~\ref{cor:complem}.
\end{proof}

\begin{example}
    Let $Q$ and $R$ as in Example~\ref{ex:int-un} and consider $F=\{P_4,P_5\}$. Let us compute the characteristic polynomial of $(Q \cap R)\setminus F$. First, compute $\mu$ and $\nu$ as the following.
    \begin{equation}
        \mu = \gcd(Y_1Y_2Y_3Y_6,Y_2Y_3Y_4Y_6, Y_3Y_4Y_5) = Y_3
    \end{equation}
    and
    \begin{equation}
\nu = (1 = Y_1)(1+Y_2)(1+Y_4)(1+Y_5)(1+Y_6)
    \end{equation}
    Finally, $(\mu\cdot \nu)(\bar{Y}) = 1$ only in $(0,0,1,0,0,0)$, i.e. $\varphi^{-1}(0,0,1,0,0,0) = \{P_3\} = (Q \cap R)\setminus F$ meaning that $\mu\cdot \nu = \xi_{(Q \cap R)\setminus F} = \xi_{Q \cap R \cap F^c}= Y_3(1+Y_1)(1+Y_2)(1+Y_4)(1+Y_5)(1+Y_6)$.
    \qed
\end{example}

Consider $\xi_A, \xi_B \in \mathscr{R}$. Define the following operations.

\begin{description}
    \item[Addition:] $\xi_A \star \xi_B = \xi_{(A \cup B) \setminus (A \cap B)}$
     \item[Multiplication:] $\xi_A \diamond \xi_B = \xi_{A \cap B}.$
\end{description}

\begin{Lemma}
$(\mathscr{R}, \star, \diamond)$ is a Boolean ring. 
\end{Lemma}

\end{document}